\documentstyle[aps,multicol,epsf,psfig]{revtex}

\newcommand{\beq}{\begin{equation}}
\newcommand{\eeq}{\end{equation}}
\newcommand{\beqa}{\begin{eqnarray}}
\newcommand{\eeqa}{\end{eqnarray}}

\begin{document}
\bibliographystyle{prsty}
\title{Quantum Distribution of Gaussian Keys with Squeezed States}

\author{N. J. Cerf,$^{1,2}$ M. L\'evy,$^1$ and G. Van Assche$^1$}
\address{$^1$ Ecole Polytechnique, CP 165, Universit\'e Libre de Bruxelles,
1050 Brussels, Belgium\\
$^2$ Jet Propulsion Laboratory, California Institute of Technology, 
Pasadena, California 91109}

\date{August 2000}
\draft
\maketitle

\begin{abstract}
A continuous key distribution scheme is proposed that relies
on a pair of canonically conjugate quantum variables. 
It allows two remote parties to share a secret Gaussian key 
by encoding it into one of the two quadrature components
of a single-mode electromagnetic field. 
The resulting quantum cryptographic information vs disturbance 
tradeoff is investigated for an individual attack 
based on the optimal continuous cloning machine.
It is shown that the information gained by the eavesdropper 
then simply equals the information lost by the receiver.
\end{abstract}

\pacs{PACS numbers: 03.67.Dd, 03.65.Bz, 42.50.-p, 89.70.+c}

\begin{multicols}{2}

Quantum cryptography---or, more precisely, quantum key distribution---is 
a technique that allows two remote parties to share a secret chain 
of random bits (a secret key) that can be used for exchanging encrypted
information \cite{benn84,benn92:prl,lutk99}.
The security of this process fundamentally relies
on the Heisenberg uncertainty principle, or on the fact that
any measurement of incompatible variables inevitably affects
the state of a quantum system. Any leak of information to an eavesdropper
necessarily induces a disturbance of the system, which is, in principle, 
detectable by the authorized receiver.

In most quantum cryptosystems proposed so far, a single photon 
(or, in practice, a weak coherent state with an average photon number 
lower than one) is used
to carry each bit of the key. Mathematically, the security is based
on the use of a pair of non-commuting observables such as the
$x$- and $z$-projections of a spin-1/2 particle, $\sigma_x$ 
and $\sigma_z$, whose eigenstates are used to encode the key.
The sender (Alice) randomly chooses to encode the key using
either $\sigma_z$ (0 is encoded as $|\uparrow\rangle$ and 
1 as $|\downarrow\rangle$)
or $\sigma_x$ (0 is encoded as $2^{-1/2}(|\uparrow\rangle+
|\downarrow\rangle)$ and 1 as $2^{-1/2}(|\uparrow\rangle-
|\downarrow\rangle)$), the choice of the basis
being disclosed only {\em after} the receiver (Bob) has measured
the photon. This guarantees that an eavesdropper (Eve) cannot
read the key without corrupting the transmission. Such a procedure,
known as BB84 \cite{benn84}, is at the heart of most of
the quantum cryptographic schemes 
that have been experimentally demonstrated in the past few years, 
which are based either on the polarization (e.~g. \cite{benn92:joc,zbin98})
or the optical phase (e.~g. \cite{town98}) of single photons. 
An alternative scheme,
realized experimentally only a year ago \cite{jenn00,naik00,titt00},
can also be used based on a pair of polarization-entangled photons instead of
single photons \cite{eker91}. It is, however, fundamentally equivalent to
BB84 (see \cite{benn92:wobell}) and it again relies on 
the algebra of spin-1/2 particles.

Recently, it has been shown that another protocol for quantum key distribution 
can be devised based on continuous variables, where 
squeezed coherent light modes are used to carry 
the key\cite{ralp00,hill00,reid99}. In these techniques, 
one exploits a pair of (continuous) canonical variables
such as the two quadratures $X_1$ and $X_2$ of the amplitude of a mode 
of the electromagnetic field, which behave just as position and momentum.
The uncertainty relation $\Delta X_1\, \Delta X_2 \ge 1/4$ then
implies than Eve cannot read both quadrature components without degrading 
the state. Even though the experimental preparation of squeezed states
is a difficult task, these schemes circumvent a main weakness of the
above-mentioned cryptosystems that is the  critical dependence 
of their security on the ability of preparing single-photon states.

In this paper, we propose an alternative squeezed-state 
quantum cryptographic scheme, which provides a means
to distribute a {\em continuous} secret key. The goal of our protocol is
to have Alice and Bob sharing a continuous key that
consists of a random list of Gaussian-distributed variables
that cannot be known to Eve. Thus, in this scenario, {\em both} the key
and the quantum variable that carries it are continuous. This is in
contrast with the schemes proposed in Ref.~\cite{ralp00,hill00,reid99}, 
which appear hybrid as a continuous quantum variable was used
to carry a discrete key element (the shared key was made of bits, or,
in general, discrete variables). Instead, our approach can be viewed 
as an {\em all-continuous} quantum cryptographic scheme, which is
the proper continuous extension of the BB84 scheme. First, from
a theoretical perspective, this provides a more satisfying continuous
treatment of quantum key distribution. Remarkably, the tradeoff 
between Eve's information gain and the disturbance at Bob's station 
can be expressed in an unexpectedly simple way (if we restrict ourselves
to an individual attack based on the optimal continuous cloning machine):
the information gained by Eve on one quadrature is at most equal
to the information lost by Bob on the other quadrature.
This results in a simple information-theoretic measure of the disturbance,
namely the defect of information at Bob's station.
Moreover, this all-continuous scheme avoids a potential attack against
the scheme proposed in Ref.~\cite{ralp00,hill00,reid99} by filling in the gaps
between the values used to encode the discrete key values (this will
be explained later on).

Let us now detail our protocol.
The uncertainty relation implies that it is impossible to measure
with full accuracy {\em both} quadratures of a single mode, $X_1$ 
and $X_2$. Alice exploits this property by 
encoding the key elements (random Gaussian samples) as 
a quadrature squeezed state either in $X_1$ or in $X_2$,
in such a way that an eavesdropper ignoring which of these
two ``bases'' is used cannot acquire information without disturbing
the state. In basis 1, Alice prepares a squeezed vacuum state such that 
the fluctuations of $X_1$ are squeezed ($\Delta X_1^2=\sigma_1^2<1/4$), 
and then applies a displacement of $X_1$ by an amount
equal to the value of the Gaussian key ($\langle X_1 \rangle=x$, where
$x$ is the encoded key element). The quantity $\sigma_1^2$ refers here
to the intrinsic variance of $X_1$ in the squeezed state; the corresponding
squeeze parameter is $r_1=-\ln(2 \sigma_1)$.
We denote by $\Sigma_1^2$ the variance
of this Gaussian key, so the mean value $\langle X_1\rangle$
is itself distributed as a Gaussian of mean 0 and variance $\Sigma_1^2$. 
Conversely, in basis 2, Alice sends a squeezed state in $X_2$
($\Delta X_2^2=\sigma_2^2<1/4$), whose displacement
encodes the Gaussian key ($\langle X_2 \rangle=x$).
Again, $\langle X_2\rangle$ has a Gaussian profile with mean 0 and
variance $\Sigma_2^2$,
while the squeeze parameter in mode 2 is $r_2=-\ln(2 \sigma_2)$.
Thus, in both basis, Alice encodes the key into a displaced vacuum squeezed
state, the squeezing (by $r$) and displacement (by $x$) being applied 
at random on $X_1$ or $X_2$.

Now, for the cryptographic setup to be secure, we require the statistical
distribution of the $X_1$ measurement outcomes to be indistinguishable
whether basis 1 or 2 is used by Alice. If this condition is fulfilled, 
Eve cannot obtain any indication on whether she is measuring 
a type 1 or type 2 squeezed state, whatever the statistics she accumulates.
If basis 1 is used, the outcomes of $X_1$ measurements (that can be
obtained in practice by homodyne detection) are distributed 
as a Gaussian of variance $\Sigma_1^2+\sigma_1^2$ since each squeezed
state gives an extra contribution of $\sigma_1^2$ to the variance.
If, on the contrary, a type 2 squeezed state is
measured, then the outcomes of $X_1$ measurements exhibit a Gaussian 
distribution of variance $1/(16 \sigma_2^2)$ as a result of the uncertainty
principle. Thus, we impose the condition
\beq
\Sigma_1^2+\sigma_1^2=1/(16 \sigma_2^2)
\eeq
Similarly, the requirement that type 1 and 2 squeezed states 
are indistinguishable when performing $X_2$ measurements implies that 
$\Sigma_2^2+\sigma_2^2=1/(16 \sigma_1^2)$. These two relations
can be summarized as
\beq  \label{eq_variances}
1+\Sigma_1^2 / \sigma_1^2
= 1+\Sigma_2^2 / \sigma_2^2
= 1 / \alpha^2
\eeq
where $\alpha=4 \, \sigma_1 \sigma_2 = e^{-(r_1+r_2)}$ 
is a (positive) dimensionless constant
which must satisfy $\alpha \leq 1$ (or $\sigma_1 \sigma_2 \leq 1/4$)
for Eq.~(\ref{eq_variances}) to be consistent. More generally,
these two conditions guarantee that the density matrices
of the encoded key elements are the same in bases 1 and 2, 
making them indistinguishable. Thus, choosing the squeeze parameters
$r_1$ and $r_2$ is sufficient to completely characterize the protocol.

Let us now analyze the transmission of the Gaussian key elements
in the case where there is no eavesdropper and the transmission
is perfect. We first need to recall some standard notions of Shannon theory
concerning the treatment of continuous transmission channels. 
Consider a discrete-time continuous channel which adds a Gaussian noise
of variance $\sigma^2$ on each signal.  
If the input $x$ of the channel is a Gaussian signal 
of variance $\Sigma^2$, the uncertainty on $x$ can be measured by 
the differential Shannon entropy
$h(x)=2^{-1} \log_2(2\pi\, {\rm e} \, \Sigma^2)$~bits\cite{cover}.
Conditionally on $x$, 
the output $y$ is distributed as a Gaussian of variance $\sigma^2$,
so that the entropy of $y$ conditionally on $x$ becomes
$h(y|x)= 2^{-1} \log_2(2\pi\, {\rm e} \, \sigma^2)$~bits. Now,
the distribution of $y$ is given by the convolution of these
two Gaussians, i.~e., a Gaussian of variance $\Sigma^2+\sigma^2$. 
Hence, the output entropy is 
$h(y) =2^{-1} \log_2(2\pi\, {\rm e} \, (\Sigma^2+\sigma^2))$~bits.
According to Shannon theory, the information that is processed through 
this noisy channel can be expressed as the mutual information
between $x$ and $y$ (the amount by which the uncertainty on $y$
is reduced by knowing $x$):
\beq  \label{eq_shannon}
I\; {\rm (bits)}=h(y)-h(y|x)= {1\over 2} \log_2\left(1+\gamma\right)
\eeq
where $\gamma=\Sigma^2 / \sigma^2$ can be viewed as the
signal-to-noise ratio (SNR).
This is Shannon's famous formula for the capacity 
of a Gaussian additive noise channel\cite{shan48}. 
Here, the signal variance (or power)
is simply $\Sigma^2$, while the noise variance is $\sigma^2$.
This capacity measures
the number of bits that can be transmitted 
asymptotically (using block coding) per use of the channel,
with an arbitrary high fidelity
for a given SNR. It can be shown to be attainable if the signal 
is Gaussian distributed (which is the case under consideration here). 

Coming back to our cryptographic setup, consider the situation 
(with no eavesdropping) where Bob performs a measurement in the
good basis after the latter is publicly announced by Alice. 
(It is equivalent to the more realistic procedure where Bob measures 
the key in a random basis, but then discards the bad outcomes 
after the basis is disclosed by Alice.)
The SNR in basis 1 is simply $\gamma_1=\Sigma_1^2/\sigma_1^2$, while
it is $\gamma_2=\Sigma_2^2/\sigma_2^2$ in basis 2.
Using this notation, Eq.~(\ref{eq_variances}) becomes
$1+\gamma_1=1+\gamma_2=1/\alpha^2$, so that we must have
the same SNR in both basis, $\gamma=e^{2(r_1+r_2)}-1$.
This means that the processed information is also the same in both bases, 
and can be expressed, using Eq.~(\ref{eq_shannon}), as  
\beq
I_0\; {\rm (bits)} = -\log_2 (\alpha) = (r_1+r_2) / \ln(2) 
\eeq
Thus, our continuous quantum cryptographic technique
can be essentially characterized by a single dimensionless constant
$\alpha$ (the product of the $X_1$ noise of type-1 squeezed states 
times the $X_2$ noise of type-2 squeezed states).
It works provided that $\alpha\leq 1$, as
a finite amount of information is then processed from Alice to Bob.
Note that $I_0$ (expressed in natural units---nats---rather
than in bits) is simply equal to the sum
of the squeeze parameters in bases 1 and 2, 
which reflects that the processed information  
is zero in the absence of squeezing, and grows linearly with squeezing
in bases 1 and 2. For example, if $\sigma_1^2=\sigma_2^2=1/8$,
i.~e., if we have a squeeze factor $e^r=\sqrt{2}$ in each basis,
then $\alpha=1/2$, so we can process one bit on average 
per use of the channel. This corresponds to $\gamma=3$ 
in both bases. More generally, we see that the processed information
in the absence of eavesdropping increases as
$\alpha$ gets smaller. In some sense, the more we violate
a pseudo-uncertainty relation $\sigma_1 \sigma_2 \geq 1/4$,
the larger this information gets. Remember that $\sigma_1$ and $\sigma_2$ 
are standard deviations of $X_1$ and $X_2$ measurements on 
type 1 and 2 states, respectively. If they referred to $X_1$ and $X_2$
measurements on a {\em same} state,  then the above uncertainty relation 
would apply, and Eq.~(\ref{eq_variances})
could not be satisfied (except for the useless case $\alpha=1$).

The average photon number contained in each encoded key state
clearly increases with the widening of the displacement ($\Sigma^2$)
needed to represent Alice's key values for a given SNR.
It also increases as squeezing increases, but then the
displacement distribution can be narrowed to achieve a same SNR. 
Let us determine the relative contribution of these two effects
focusing on one basis, and assuming for simplicity that 
$\sigma_1 = \sigma_2 = \sigma$
so that the same squeezing is applied on both quadratures. In
this case, Eq.~(\ref{eq_variances}) implies
that $\sigma^2=\frac{1}{4} e^{-2r}$,
$\Sigma^2= \frac{1}{2} \sinh(2r)$, and $1+\gamma=e^{4r}$.
For a given encoded key state (with a squeeze parameter $r$
and displacement $x$, where $x$ is the key value Alice wishes to transmit), 
the mean photon number can be written as
$N= x^2 + \sinh^2 r$, where the first term reflects
the displacement effect while the second 
characterizes vacuum squeezing\cite{scully}.
For a given SNR $\gamma$ (or a given squeezing
parameter $r$), we obtain the average number of photons 
over all possible values $x$ sent by Alice (distributed as a Gaussian 
of mean 0 and variance $\Sigma^2$), 
$ \langle N \rangle = \Sigma^2 + \sinh^2 r$.
Using the relation between $\gamma$ and $r$, this gives for the average
number of photons per key pulse:
\beq 
\langle N \rangle = {1-\alpha \over 2\alpha } = {e^{2r}-1 \over 2} 
= {(1+\gamma)^{1/2}-1 \over 2} 
\eeq
Equivalently, the processed information can be expressed
as a function of the average photon number,
\beq
I_0\; {\rm (bits)}=\log_2(2\, \langle N \rangle +1)
\eeq
implying that the photon number must increase exponentially
with the processed information.

We shall now investigate the tradeoff between the information acquired
by Bob and Eve in this continuous cryptographic protocol. First, we should
emphasize that, even in the absence of eavesdropping, the key elements
received by Bob are not exactly equal to those sent by Alice. This is
in contrast with BB84, and is simply due to the fact that 
the noise due to the intrinsic fluctuations of the squeezed states 
always adds to the signal, giving
rise to a finite SNR. This already holds at Alice's station, regardless
the (possibly tapped) channel. So, an eavesdropper will be visible
in this scheme by an enhanced noise variance (or a reduced SNR) at Bob's
station. A protocol that Alice can follow to detect any eavesdropping
can be to disclose, on a public channel, the exact values $x$ of 
a random subset of key elements. Then, Bob compares them
to the received values $y$ and computes the distribution of the
differences $y-x$. For a perfect and untapped channel, it should
be a Gaussian of variance $\sigma^2$, so the SNR is unchanged.
Otherwise, the SNR decreases by an amount that can be viewed
as a measure of the disturbance of the Alice-to-Bob channel.
Assume, for example, that Eve uses an individual ``intercept-and-resend''
attack, measuring each key element in basis 1 or 2, at random,
and resending a squeezed state centered on the value 
of the measured quadrature. The variance at Bob's station 
is $2\sigma^2$ (twice the intrinsic variance!)
if Eve used the good basis,
or $1/(16\sigma^2)$ in the opposite case, so the resulting noise
variance is $\sigma^2 [1+1/(2\alpha^2)]$. Thus Bob's computed SNR 
is reduced by a factor $2/(3+\gamma)$.

Let us now make the assumption that the optimal individual eavesdropping
strategy for Eve consists in using the optimal (Gaussian) cloning machine
for continuous quantum variables\cite{cerf00:cont,cerf00:coherent}.
This is a very sensible conjecture as the phase-covariant qubit cloner
is known to be the best individual eavesdropping strategy 
for BB84\cite{fuch97} (actually, the universal qubit cloner is optimal
for the related six-state quantum cryptographic 
protocol\cite{brus98,bech99:6state}).
We consider an attack where Eve makes two imperfect copies of the
key element, and sends one of them to Bob while she keeps the other one. 
Bob and Eve both wait until Alice reveals the basis
she used for encoding the key before measuring the received state
in the appropriate basis (again, this is equivalent to Bob measuring
in a random basis and then discarding the bad measurements after
the basis disclosure). To analyze the information-theoretic balance
between Bob and Eve, we use a general class of {\em asymmetric}
Gaussian cloners defined in Ref.~\cite{cerf00:cont} that
result in a different amount of noise on both quadratures
and for Bob and Eve. It is proven in Ref.~\cite{cerf00:cont}
that the inequality
\beq
\sigma_{1,B}^2 \, \sigma_{2,E}^2 \geq 1/16
\eeq
must hold (and is saturated for this class of cloners), 
where $\sigma_{1,B}^2$ and $\sigma_{2,E}^2$ are
the variances of the errors that affect
Bob's $X_1$ measurements and Eve's $X_2$ measurements, respectively.
For example, if basis 1 is used, then the outcomes of $X_1$ measurements
on Bob's side will be distributed as a Gaussian 
of variance $\sigma_1^2 + \sigma_{1,B}^2$
since cloning-induced errors are superimposed on
the intrinsic fluctuations of the squeezed states.
Similarly, a second no-cloning uncertainty relation holds,
connecting Bob's errors on $X_2$ and Eve's errors on $X_1$:
$\sigma_{2,B}^2 \, \sigma_{1,E}^2 \geq 1/16$.
Let us now characterize the cloners that saturate these
inequalities by two parameters $\chi$ and $\gamma$: 
we rewrite the error variances on Bob's side as
$\sigma_{1,B}^2=\chi \gamma (\sigma_1^2/\alpha)$ and
$\sigma_{2,B}^2=\chi \gamma^{-1} (\sigma_2^2/\alpha)$,
while the errors on Eve's side are written as
$\sigma_{1,E}^2=\chi^{-1}\gamma (\sigma_1^2/\alpha)$
and $\sigma_{2,E}^2=\chi^{-1} \gamma^{-1} (\sigma_1^2/\alpha)$.
Thus, $\chi$ characterizes the balance between Bob's and Eve's errors 
as $\sigma_{1,B}/\sigma_{1,E}=\sigma_{2,B}/\sigma_{2,E}=\chi$.
The limit $\chi\to 0$ corresponds to the case where
Bob has a negligible cloning-induced additional error 
on his measured quadratures, so he
gets the entire information $I_0$ (Eve does not get any information). The
case $\chi=1$ represents a symmetric situation where the errors
induced by cloning are the same for Bob and Eve. Of course, the limit
$\chi\to\infty$ is the opposite situation where Eve gets most 
of the information with no error. 
Similarly, $\gamma$ describes the quadrature 1 vs 2 balance,
that is, $\sigma_{1,B}/\sigma_{2,B}=\sigma_{1,E}/\sigma_{2,E}=
\gamma (\sigma_1/\sigma_2)$.

Now, we need to express the information processed from Alice to Bob
(or from Alice to Eve) in basis 1 (or basis 2).
In basis 1, the variance of Bob's measurement outcomes is
$\sigma_1^2+\sigma_{1,B}^2=(1+\chi\gamma/\alpha)\sigma_1^2$, 
while the distribution of the key
elements has a variance $\Sigma_1^2$. Using Shannon's formula, 
Eq.~(\ref{eq_shannon}), and the identity $1+\Sigma_1^2/\sigma_1^2=1/\alpha^2$,
we obtain the information processed from Alice to Bob
in basis 1:
\beq  \label{eq_IAB}
I_{1,B}= {1\over 2} \log_2
\left( {1+\alpha\chi\gamma \over \alpha^2 + \alpha\chi\gamma} \right)
\eeq
Similarly, using the variance of Eve's outcomes in basis~2,
$\sigma_2^2+\sigma_{2,E}^2=[1+1/(\chi\gamma\alpha)]\sigma_2^2$, 
an analogous calculation yields for Eve's information in basis 2 
\beq   \label{eq_IAE}
I_{2,E}= {1\over 2} \log_2
\left( {1+\alpha/(\chi\gamma) \over \alpha^2 + \alpha/(\chi\gamma)} \right)
\eeq
Finally, the balance between Bob's and Eve's information can be
expressed by calculating the {\em sum} of equations
(\ref{eq_IAB}) and (\ref{eq_IAE}):
\beq  \label{eq_sum}
I_{1,B}+I_{2,E}={1\over 2} \log_2(1/\alpha^2) = I_0
\eeq
Remarkably, it appears that the information acquired
by Eve on the second quadrature, $I_{2,E}$, is {\em exactly} counterbalanced
by the defect of information at Bob's side on the first quadrature, 
$I_0-I_{1,B}$. Of course, the counterpart of Eq.~(\ref{eq_sum}) also holds 
when interchanging the bases, that is, $I_{2,B}+I_{1,E}= I_0$.

Thus, assuming that the use of the continuous cloner
is the best possible individual attack against our
continuous cryptographic protocol, Bob's information
loss can be viewed as a proper disturbance measure
as it simply is an upper bound on the information that might 
potentially have been gained by an eavesdropper.
Consequently, the net amount of key bits that can be generated 
by this method is bounded by $I_{B}-I_{E}=I_0 - 2 I_{E}$. 
This follows from
\cite{maur93} where it is proven that the secret key rate of $A$ and $B$
with respect to $E$ is lower bounded by the difference of mutual information
$I(A;B)-I(A;E)$. Even though
$A$, $B$ and $E$ here denote continuous variables, we can use this
result provided that the generated key and the exchanged 
reconciliation messages are discrete as required in \cite{maur93}. 
Our continuous variables $A$, $B$ and $E$ 
only appear at the right of the conditional bar in entropy
formulas, so they can be approximated by discrete numbers
(that is, they can be replaced by an integer
such as $\lfloor n A \rfloor$, approximating the real variable $A$). As
$n$ grows, it will soon be close to the real variable with a
precision far beyond what is needed given the noise level. 
Thus, we conclude that our protocol can
only work provided that $I_{E} < I_0 /2$, that is, iff $\chi < 1$.
Stated otherwise, the quality of the signals measured by
Alice and Bob must be bounded by $I_{B} > I_0/2$, or in terms of
signal-to-noise ratios $\gamma' > \sqrt{1+\gamma}-1$, where $\gamma'$ is
the SNR measured by Bob. 
This means that a 1-bit channel ($\gamma = 3$) may
still be used if the noise power is almost tripled ($\gamma' > 1$).
In summary, the procedure we propose here consists in the quantum distribution
of a (real) Gaussian key, followed by a discretization procedure
so as to apply some (discrete) reconciliation and privacy amplification 
protocol. Such a strategy avoids a weakness of the squeezed-state
cryptosystems as presented in Refs. \cite{ralp00,hill00,reid99}.
There, the key is binary (or belong to a larger finite alphabet), so
there are always gaps between the discrete key values. This
allows Eve to gain knowledge about the occurrences where she measured 
the wrong quadrature (without getting the key value). This knowledge
alone is sufficient for her to attack this key distribution scheme 
simply by omitting to resend the corresponding key elements to Bob, 
thereby faking a small attenuation in the transmission.
This limitation does not apply to our scheme since the continuous
key values fill in an entire region in the $(X_1,X_2)$ phase space.


In conclusion, an all-continuous quantum cryptographic protocol was proposed  
that is based on single-mode squeezed states of the electromagnetic
field. It exploits the uncertainty relation between
the conjugate pair of quadrature components $X_1$ and $X_2$
by encoding a continuous Gaussian-distributed key 
into either $X_1$- or $X_2$-squeezed states, thereby allowing a continuous
key distribution between two remote parties. It is shown that
the information acquired by an eavesdropper on the key elements
encoded in $X_1$ is compensated by a reduction (by a same amount) 
of the key information available on the $X_2$ amplitude at the receiver's
station. This information-theoretic tradeoff characterizes the
worst-case individual attack based on the cloning machine, so 
we conclude that the loss of information at the receiver's end
is a good upper bound on the tapped information.
A realization of this continuous protocol based on squeezed states
would be very challenging, as the generation of squeezed light
has been a difficult experimental target for years. Also, it
would require synchronized local oscillators at Alice's and Bob's stations,
in order for them to have a common phase for homodyne detecting
the amplitudes $X_1$ and $X_2$. In addition, probably the main 
limitation in the implementation of this protocol is related
to the loss of squeezing effected by attenuation in the transmission medium.
This would dramatically decrease the SNR, and make the protocol 
less efficient (or insecure). In analogy with what is known for BB84,
there probably is a threshold on the squeeze parameter 
that Alice should reach, below which the protocol would fail.
Nevertheless, it should be stressed that the cryptographic protocol
proposed here was analyzed using the conjugate pair $X_1$ and $X_2$,
but other complementary variables might be exploited as well.
In particular, one could possibly imagine a continuous cryptographic
scheme based on the time-frequency complementarity, where
ultra-short single-photon pulses, or, alternatively,
single-photon pulses that are highly resolved in frequency would
be used in order to encode the Gaussian key. Such a scheme
might possibly avoid some of the weaknesses of the squeezed state
protocol, and be more appropriate for an experimental realization.


We are grateful to Jonathan Dowling, Nicolas Gisin, Serge Massar, and
Hugo Zbinden for helpful discussions. G. V. A. acknowledges support
from the Banque Nationale de Belgique.

\end{multicols}
\end{document}